# Effects of intrinsic strain on the structural stability and mechanical properties of phosphorene nanotubes

Xiangbiao Liao[1], Feng Hao[1], Hang Xiao[1], and Xi Chen[1,2,*]

[1]Department of Earth and Environmental Engineering, Columbia University, New York, NY 10027, USA

[2]International Center for Applied Mechanics, SV Laboratory, School of Aerospace, Xi'an Jiaotong University, Xi'an 710049, China

[*]Corresponding author: xichen@columbia.edu

**Abstract**

Using molecular dynamics (MD) simulations, we explore the structural stability and mechanical integrity of phosphorene nanotubes (PNTs), where the intrinsic strain in the tubular PNT structure plays an important role. It is proposed that the atomic structure of larger-diameter armchair PNTs (armPNTs) can remain stable at higher temperature, but the high intrinsic strain in the hoop direction renders zigzag PNTs (zigPNTs) less favorable. The mechanical properties of PNTs, including the Young's modulus and fracture strength, are sensitive to the diameter, showing a size dependence. A simple model is proposed to express the Young's modulus as a function of the intrinsic axial strain which in turns depends on the diameter of PNTs. In addition, the compressive buckling of armPNTs is length-dependent, whose instability modes transit from column buckling to shell buckling are observed as the ratio of diameter/length increases.

**Introduction**

Phosphorene is a monolayer of black phosphorus (BP). The layered structure possessing the strong intralayer strength and weaker interlayer interaction like graphite [1]. The novel two-dimensional (2D) functional phosphorene has become the focus of significant research effort recently, thanks to the successful fabrications by micromechanical cleavage [2], Ar$^+$ plasma thinning process [3] and liquid exfoliation methods [4]. Not only does phosphorene exhibit comparably high carrier mobility (~10000 cm$^2$V$^{-1}$s$^{-1}$), it is also a semiconductor with band gap (0.3–2 eV) [5] larger than semimetallic graphene [6]. Additionally, the puckered structure of phosphorene, shown in Fig. 1 (a), enables its significant anisotropy of band dispersion [7], electrical and thermal conductivity [7-9], mechanical properties [10-12]. These properties suggest extensive potential applications in phosphorene-based nanodevices, including transistors, advanced batteries and optoelectronics [13,14].

Despite these advances, limited attention has been paid thus far to the phosphorus allotrope, the phosphorene nanotubes. PNTs were theoretically designed and predicted by rolling up a phosphorene sheet along armchair or zigzag direction, forming two types of nanotubes, ($m$, 0) zigPNTs and (0, $n$) armPNTs shown in Figs. 1(b) and 1(c), respectively [15,16]. The configuration of a PNT was characterized by the integer indexes ($m$, $n$). Further, a wrapping vector $\boldsymbol{R} = m\boldsymbol{a_1} + n\boldsymbol{a_2}$ has been defined [17,18]. Indeed, PNTs' anisotropies in energetic viability and phase transition [15,19,20] were assessed and predicted through first principle calculations, leading to the possibility of synthesis of α-PNTs and β-armPNTs wrapping from blue and black phosphorene [20], respectively. However, the β-zigPNT structure was regarded as unfavorable due to large strain energy [20]. Furthermore, diagrams in determining stable, faceted PNTs and fullerene structures were presented [19] and defect-induced blue PNTs with neglected bending energy were demonstrated to have lower formation energy than round PNTs by density functional theory [21]. Although these theoretical predictions at 0 K offered a guidance for future laboratory fabrications, evaluation of the stability of PNTs at finite temperature remains unrealized.

Strain-engineering of PNTs was demonstrated possible. For example, strain could affect the carrier mobility and band structures of PNTs [17,20,22], and the elastic modulus and conductance [22] could be varied by its diameter. No significant difference of optical properties was found by varying the diameter of PNTs, though the chirality and polarization direction dependences were presented [20]. Compared with the phosphorus monolayer, PNTs are more favorable to become practical structures in nanodevice applications, such as strain sensors, photodetectors, and transistors due to the great tunability of electrical and optical properties by size and strain [22]. For the aforementioned strain-engineered applications, the mechanical behaviors of PNTs, including Young's modulus, facture strength and buckling strain, require further investigation.

In this work, MD simulations are carried out to study the thermal-stability and mechanical behaviors of armPNTs and zigPNTs. Intriguingly, PNTs with larger diameter are likely to withstand higher temperature due to the relatively lower intrinsic hoop strain, and the armPNTs can resist higher thermal load than zigPNTs at the same size. Based on the stable armPNT and zigPNT structures, the size-dependence of Young's modulus and fracture strength are observed. To clarify the underlying mechanism, an analytical continuum model is developed to illustrate the size dependence of the Young's modulus. Finally, upon axial compression, the buckling mode transition of armPNTs from column buckling to shell buckling is observed with increasing tube diameter, which may shed some light for strain-tunable characteristics and operation of future PNTs-based electronics.

**Model and Method**

Large-scale Atomic/Molecular Massively Parallel Simulator (LAMMPS)-based MD simulations [23] are performed. The Stillinger-Weber (SW) potential [24] validated to describe the mechanical properties of phosphorene is used in the simulation. The boundary in the axial direction of PNT is periodical, while a sufficiently large vacuum space (100 Å) surrounding the nanotube is applied. For the stability analysis, each initial PNT, having the length of 50 supercells, is relaxed to a thermally stable state with a NPT ensemble at given temperature (0 K~400 K), controlled by the Nose-Hoovers thermostat for 250 ps. The pressure and time step are set at 0 bar and 0.5 fs, respectively. The above equilibrium structures at 0 K are used to study the deformation behaviors. The mechanical properties under axial tension and compression are studied at a constant strain rate of $10^{-4}\,ps^{-1}$ in the NPT ensemble. The strain is defined as the relative change of simulation box along the axial direction ($\varepsilon=\Delta L/L_z$). In order to calculate the stress, the interlayer spacing of phosphorene is taken as the thickness of a PNT, commonly assumed to be 5.24 Å [12]. For the sake of simplicity, only zigPNTs and armPNTs (Figs. 1(b) and 1(c)) are studied, and other mixed chiralities will be explored in future.

**Results and Discussion**

1. **Thermal stability of PNTs.**

The strain energy $E_s$, originating from bending a phosphorene sheet into a PNT is defined as

$$E_s = \frac{E_{PNT} - E_{BP}}{N}$$

where $N$ is the total numbers of atoms in the simulated PNT, and $E_{PNT} - E_{BP}$ is the potential energy difference in an equilibrium nanotube with respect to the 2D phosphorene sheet (as the reference system) at a given temperature. Physically, a higher strain energy for nanotube means a larger intrinsic strain in the hoop direction.

The results in Figs. 2(a) and 2(b) provide maps for determining the stable region by varying temperature and size. All single-walled PNTs are inclined to have higher thermal stability as the diameter increases. Taking the (0, 10) armPNT and (0, 20) armPNT as examples, the former is able to resist $T$ = 175 K which is lower than that of the later one, $T$ = 410 K. Figs. 2(c) and 2(d) explain the cause of this phenomenon. It is evident that a higher strain energy is stored in a PNT with smaller diameter under higher thermal load after energy minimization. The results at 0 K agree well with those from the first principle calculations [18,20]. Thus, the higher intrinsic hoop strain in a bended PNT structure is responsible for collapsing at lower thermal load. In addition, the strain energies in the (0, 6) armPNT (0.085 eV/atom) and (17, 0) zigPNT (0.093 eV/atom) are close to the value at fracture (~0.1 eV/atom) for a 2D phosphorene sheet under uniaxial tension along both armchair and zigzag directions, indicating the verge of integrity. Hereby, the (0, 6) armPNT and (17, 0) zigPNT are suggested to be the smallest stable PNTs without fracture or phase trainsition [18], which echoes with the collapsed smaller structures found in our MD simulation.

In terms of the effect of chirality, at a particular radius, armPNTs are found to have the ability to withstand much higher thermal loads than zigPNTs, by comparing Fig. 2(a) with Fig. 2(b). This is exemplified by the maximum temperature, $T$ = 410 K, upon which the (0, 20) armPNT could resist, as opposed to that of the (20, 0) zigPNT which may only resist $T$ = 10 K. By comparing the strain energies of the two types of PNTs, the values for armPNTs are much lower than that for zigPNTs at all temperatures and sizes shown in Figs. 2(c) and 2(d). The structural stability upon different thermal loads provides a preliminary guideline for future synthesis in laboratory and operation of PNTs in applications. Note that some other phases are not considered in the present manuscript, such as the faceted PNTs with joints [16,21] and the ones with bending-induced phase transitions [18]. The investigation of their thermal stabilities will be subjected to future research.

## 2. Size-dependent tensile properties

To further explore the underlying mechanical properties of PNTs, initially stable PNT structures are stretched at 0 K without involving thermal fluctuations. Figs. 3(a) and 3(b) show the nominal stress-aixal tensile strain curves of (0, $n$) armPNTs and ($m$, 0) zigPNTs, respectively. ArmPNTs (~94 GPa) are stiffer than zigPNTs (~20 GPa), while zigPNTs have larger fracture strain, which originates from the structural anisotropy of phosphorene [12,25]. Similar brittle fracture behaviors of carbon nanotubes (CNTs) at low temperature were also observed in PNTs [26]. However, PNTs are much softer than other 1D materials, for example, CNTs having the higher Young modulus (~1.0 TPa) and tensile strength (~140 GPa) [27].

It is evident that the wrapping index or diameter has a significant influence on the stress-strain curve, showing strong size effect on both the stiffness and strength of the PNTs. The Young's moduli of PNTs in Figs. 4(a) and 4(b) can be deduced by fitting the stress-strain curves in the small strain region ($\leq 1.0\%$). As the tube diameter increases, the Young's modulus increases from 84 GPa to 102 GPa for armPNTs and from 15.6 GPa and 24 GPa for zigPNTs. Meanwhile, the effects of diameter on the tensile strength of PNTs are shown in Figs. 4(c) and 4(d). The present values of the fracture strength for both armPNTs and zigPNTs increase with increasing nanotube diameter, approaching the limit values of 10.3 GPa and 4.0 GPa for a pristine phosphorene sheet [24].

It is interesting to find the existence of bending-induced intrinsic stress along the axial direction in equilibrium PNTs, even before external loads are applied (shown in Fig. 5(a)). The residual compressive stress in the inner sublayer of PNTs is relaxed when the applied strain increases up to $\varepsilon_0$, which is defined as the initial axial strain. The unique atomic structure (two-sublayers phosphorous atoms shown in Fig. 5(a)) differs from monolayer atoms in CNTs [28], and is likely responsible for the intrinsic axial strain. According to Fig. 5(b), the absolute values of intrinsic axial strains for both armPNTs and zigPNTs exponentially decrease as radius increases, approaching zero. Consequently, for the stretched PNT with smaller diameter, the lower fracture strength (Figs. 4(c) and 4(d)) is caused by the larger intrinsic strain in the axial direction.

In order to explain the size-dependence of Young's modulus, a simple continnum model is proposed by considering both the intrinsic axial strain and the nonlinear effect of potential. As a first order approach, several simplified assumptions are adopted. First, despite the two sublayers (shown in Fig. 5(a)) in a PNT with bond interaction in between, the continuum model considers only a monolayer tube with a uniform intrinsic axial strain; Second, the effect of curvature on the atomic potential is neglected, although geometry nonlinearity was regarded as being responsible for the size-dependence of stiffness of CNT with diameters less than 0.5 nm [29], the PNTs in the present study are sufficiently large such that the curvature effect may be small. The nonlinear effect of potential is illustrated by the stress-strain curves of a 2D phosphorene sheet under tension at 0 K, presented in Figs. 6(a) and 6(b). This nonlinearity suggests that the Young's modulus of phosphorene is also strain-dependent, and the corresponding $E(\varepsilon)$ is plotted in Fig. 6(c) based on the fitting of a simple quadratic function. The result closely matches that calculated by Jiang *et al.* [11].

To take into account the effect of the intrinsic axial strain ($\varepsilon_0$), the Young's modulus of a continuum tube with intrinsic strain can be derived as

$$E^* = \frac{E(\varepsilon_0) + E(-\varepsilon_0)}{2} - \frac{(E(-\varepsilon_0) - E(\varepsilon_0))\varepsilon_0}{2\varepsilon_{app}} \qquad (1)$$

where $\varepsilon_{app}$ is the applied strain to deduce Young's modulus, which is sufficiently small region so that the Young's modulus of each sublayer can be regarded as a constant during deduction. For example, $\varepsilon_{app}$ is taken as 0.01 in the MD simulation above. Both the magnitude of $\varepsilon_0$ and the nonlinear variation of stiffness with respect to the strain, contribute to the size effect: as the diameter of PNTs increases, Fig. 5(b) shows the decreased intrinsic axial strain $\varepsilon_0$. While the first term in Eq. (1) remains constant due to the assumed linearity of the Young's modulus-strain relation (see Fig. 6(c)), the second decreases based on the Fig. 6(c), which consequently leads to the increased Young's modulus $E^*$. The trends given by this model are in qualitative agreement with the size-dependence of Young's modulus derived by MD simulations (see Figs. 4(a) and 4(b)).

Additionally, it is seen that the size-induced stiffening for armPNTs is more sensitive than that for zigPNTs. Basically, the nonlinear effect in zigzag direction is more significant than that in armchair direction as shown in Fig. 6(c). Specifically, as the strain increases from -0.01 to +0.01, the strain-induced reduction of Young's modulus in zigzag direction (30%) is larger than that in armchair direction (10%). This primary mechanism for size-dependence of Young's modulus is different from that for CNTs, where no intrinsic axial strain exists [29,30].

### 3. Buckling behavior of armPNTs

Compared with armPNTs, as candidate semiconductors for promising applications in field effect transistor and nanodevices, zigPNTs were reported to possess complexity in several semimetal properties [22]. Furthermore, the stability analysis shows that armPNTs having large size-range of stable structures allow the high tunability of electronic propterties at room temperature. This motivates our examination of the axial buckling behaviors for armPNTs. The relation between the critical strain of buckling, defined as the point where the first sudden drop of stress shown in the inset of Fig. 7(a), and length of (0, 10) armPNTs is plotted in Fig. 7(a). It is evident that the critical load is sensitive to length (*L*). The nanotube fails by column buckling at large slenderness ratio (*L/D*), but by squashing at small slenderness ratio when applied load reaches the ultimate compressive strain of PNTs. A similar transition was also found in the buckling behavior of a CNT [31-33]. The column buckling strain simulated by MD is in qualitative agreement with that predicted by Euler theory for an orthotropic elastic thin wall beam [34],

$$\varepsilon_{column} = \frac{2\pi^2 R^2}{L^2}(1+\frac{2}{3}\frac{\pi^2 R^2}{L^2}\frac{E_x}{G_{xy}})^{-1} \qquad (2)$$

where $R = 7.16$ Å is the radius for (0, 10) armPNT, and the Young's modulus and shear modulus of a 2D phosphorene sheet are employed here, i.e. $E_x = 105$ GPa and $G_{xy} = 25.4$ GPa [25]. The differences in mechanical parameters between PNTs and phoshorene are neglected. Large intrinsic strain (~0.5%), boundary effects, and the finite thickness of the nanotube make contributions to the discrepancy between the two predictions.

The diameter-sensitive critical strains are also shown in Fig. 7(b). As the diameter increases, the buckling mode of armPNTs transfers from column buckling to shell buckling. This transition is similar to the compressive behaviors of CNTs previously reported [35,36]. On the other hand, based on Donnell theory for orthotropic elastic thin shells [34], the critical strain is given by

$$\varepsilon_{cr} = \sqrt{\frac{2\sqrt{E_y/E_x}\,G_{xy}/E_x}{3(1-\sqrt{\nu_{xy}\nu_{yx}})}} \frac{t}{R} \quad (3)$$

where the thickness of armPNTs is assumed to be 5.24 Å, the Young's modulus along armchair direction $E_y$ = 23 GPa [25], $\nu_{xy}$ = 0.93 and $\nu_{yx}$ = 0.40 are the Poisson ratios in zigzag direction and armchair direction, respectively [11]. The results from Donnell theory agree well with those from MD simulations when the diameter of nanotube is far from the transition region (10 Å< $r$ <18 Å). Besides the large intrinsic strain mentioned above, the diameter-dependence and nonlinear response of mechanical properties, such as Young's modulus, are hardly captured by the continuum model, which may be responsible for the deviation between the two predictions. Other possible contributions to the error of Eq. (3) including the finite thickness of armPNT wall [34] and the uncertainty of its value [30], as well as imperfection in atomic coordinates (defects) which may serve to reduce the buckling strain. An improved and comprehensive model which incorporates these factors will be subjected to future study.

**Conclusion**

We carry out MD simulations on single-wall PNTs to study the structural stability and mechanical behaviors of PNTs. The results indicate that PNTs with a larger diameter are able to resist higher temperature. However, zigPNTs have lower resistance to the thermal loads due to high intrinsic hoop strain in the the wrapped structures. Also, the remarkable size-effect of mechanical properties is revealed, and both Young's modulus and fracture strength of PNTs decrease as the diameter decreases. In addition, a continuum model is developed to uncover the role of the intrinsic axial strain on the size-dependence. Finally, buckling behaviors of armPNTs are explored. It is demonstrated that the failure mode of armPNTs with a small diameter transfers from column buckling to squashing as the length decreases, and buckling mode transforms from column buckling to shell buckling when the diameter of nanotube increases. The continuum theories are in qualitative agreement with MD simulations. More quantitative modeling effort will be carried out in future. Since PNTs' excellent properties in electrics and thermoelectrics enable them as promosing components, the present study on the basic mechanical behaviors of PNTs may offer a guidance for the fabrication and strain engineering of PNTs-based on nanodevices.

**Acknowledgements**

X.C. and F. H acknowledge the support from the National Natural Science Foundation of China (11172231 and 11372241), ARPA-E (DE-AR0000396) and AFOSR (FA9550-12-1-0159); X.L. and H.X. acknowledge the China Scholarship Council for the financial support.

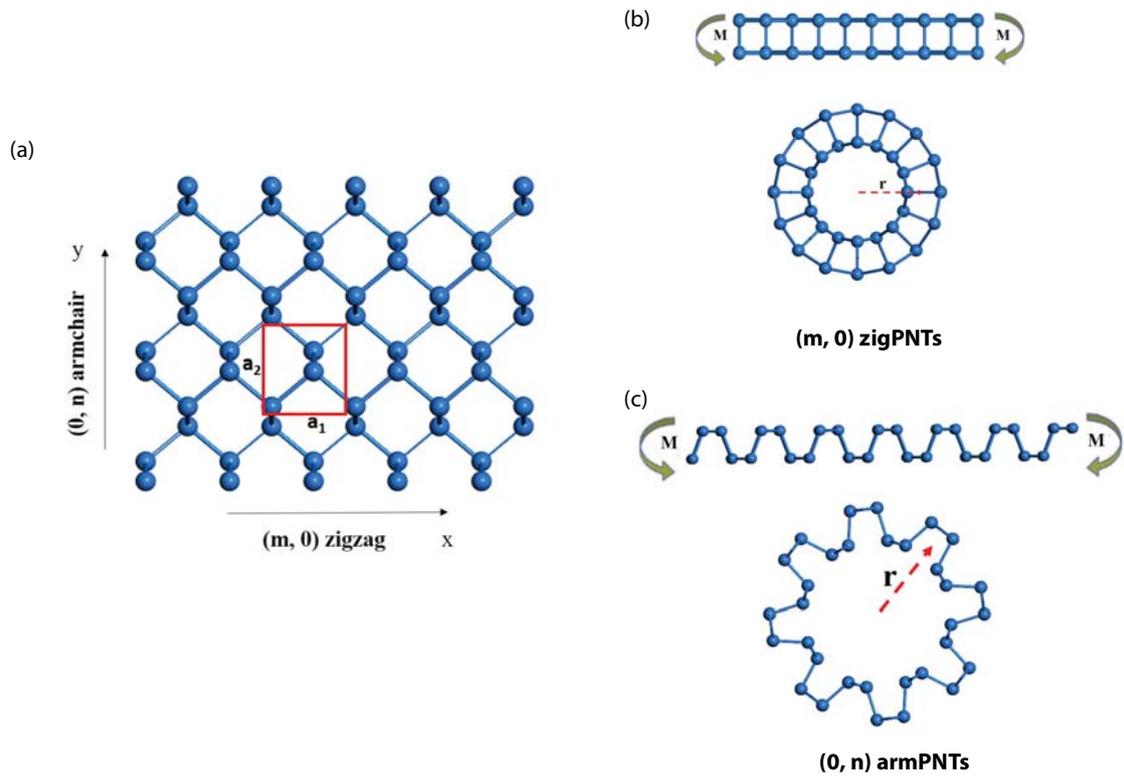

**Figure 1.** (a) Top view: the optimized structure of 2D phosphorene, and a unit cell is shown inside the red box with vectors $a_1$ = 3.3 Å and $a_2$ = 4.5 Å along the zigzag and armchair direction, respectively. (b) The (m, 0) zigPNT is bended from phosphorene sheet along the wrapping vector (m, 0), and its radius can be calculated as $r = ma_1/2\pi$. (c) the (0, $n$) armPNT is bended from MBP along the wrapping vector ($m$, 0), and its radius can be calculated as $r = na_2/2\pi$.

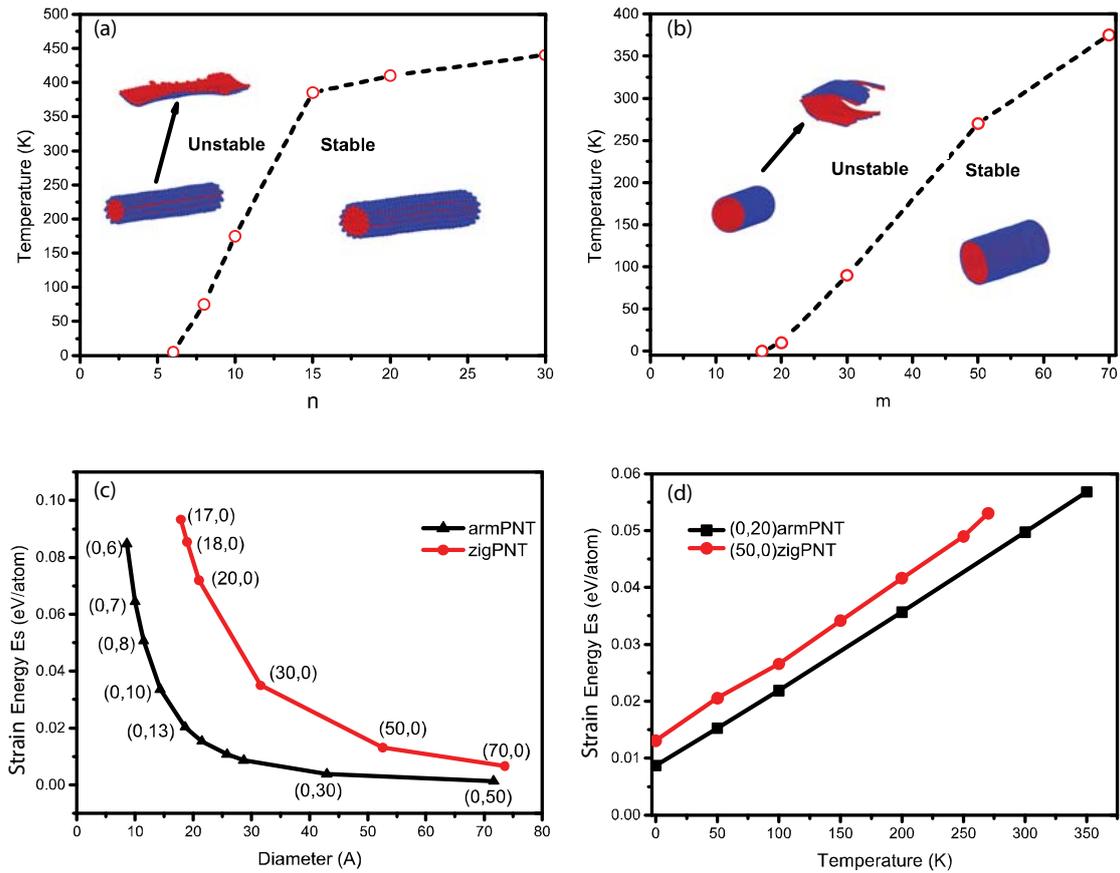

**Figure 2.** The phase diagrams for thermal-stability of (a) armPNTs and (b) zigzagPNTs with varying temperature and wrapping vector of nanotube. Stable and unstable atomic structures are shown. (c) The strain energy stored in wrapped PNTs as function of the diameter of armPNTs (black triangular) and zigPNTs (red circles) at $T = 0$ K. (d) The strain energies of (50, 0) zigPNT (red) and (0, 20) armPNT (black) change with temperature.

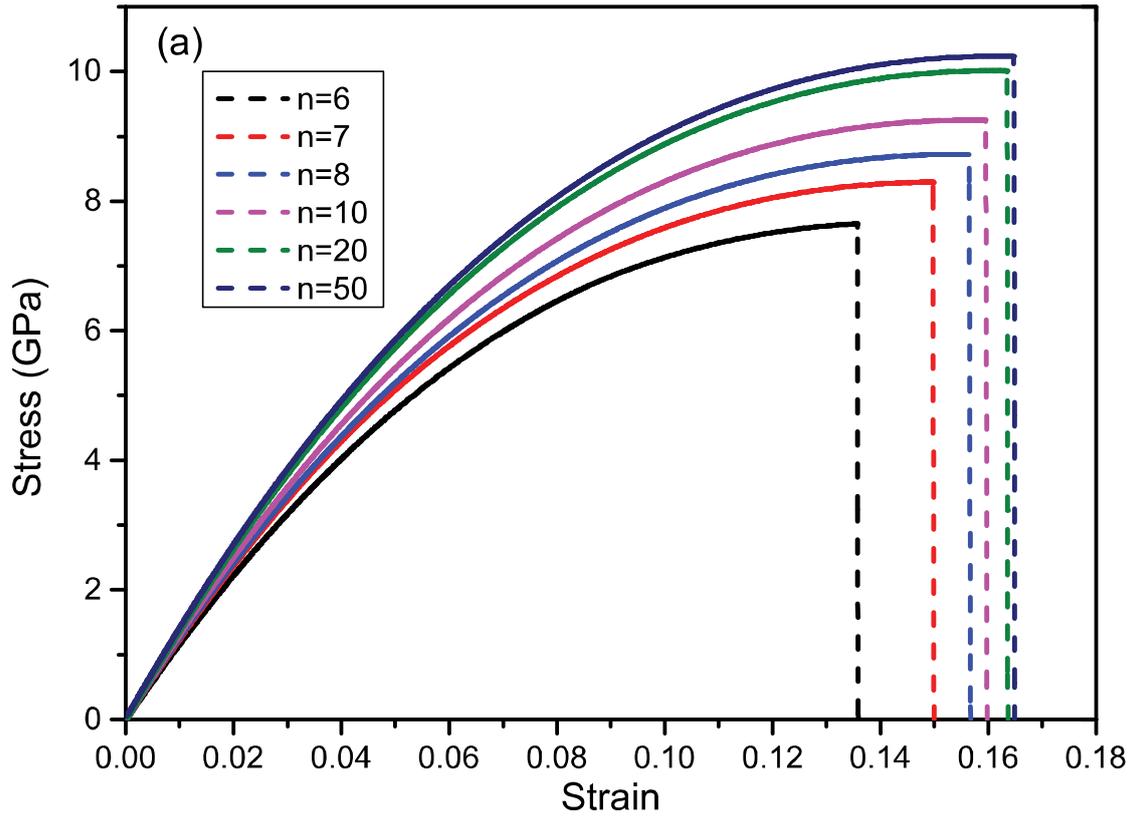
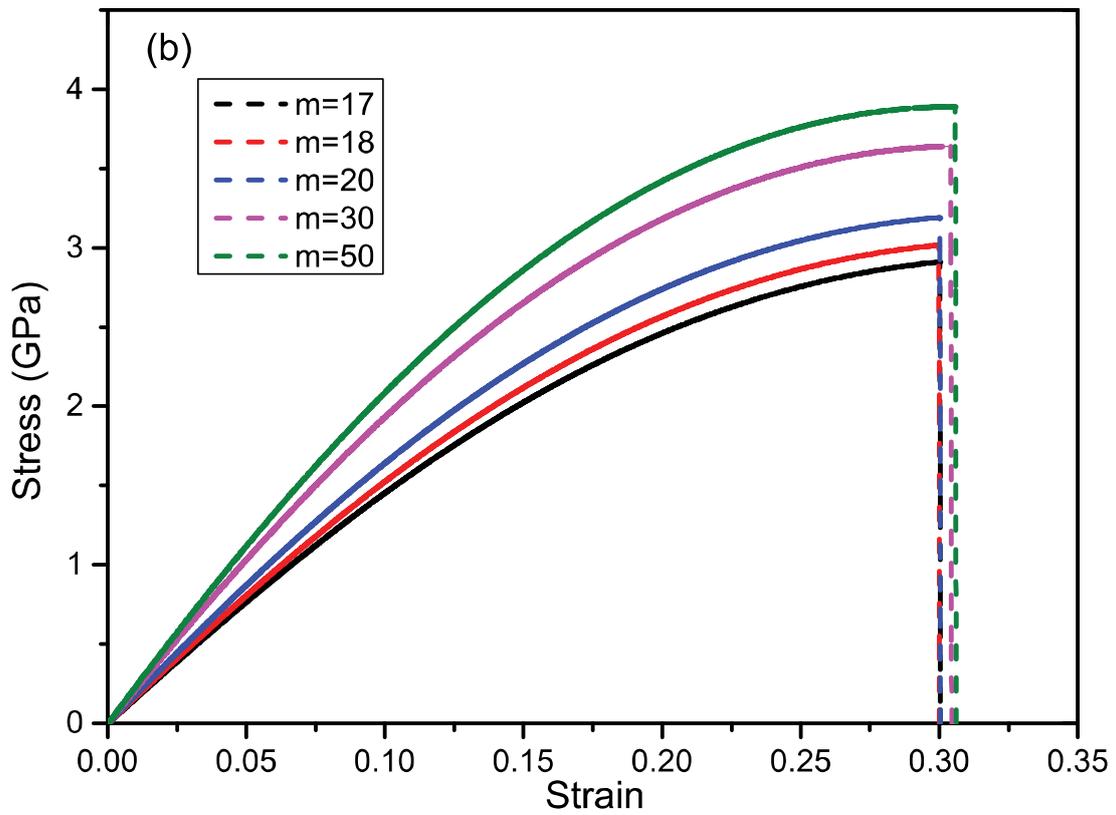

**Figure 3.** The axial tensile stress-strain curves for (a) armPNTs and (b) zigPNTs with various wrapping indexes at $T = 0$ K

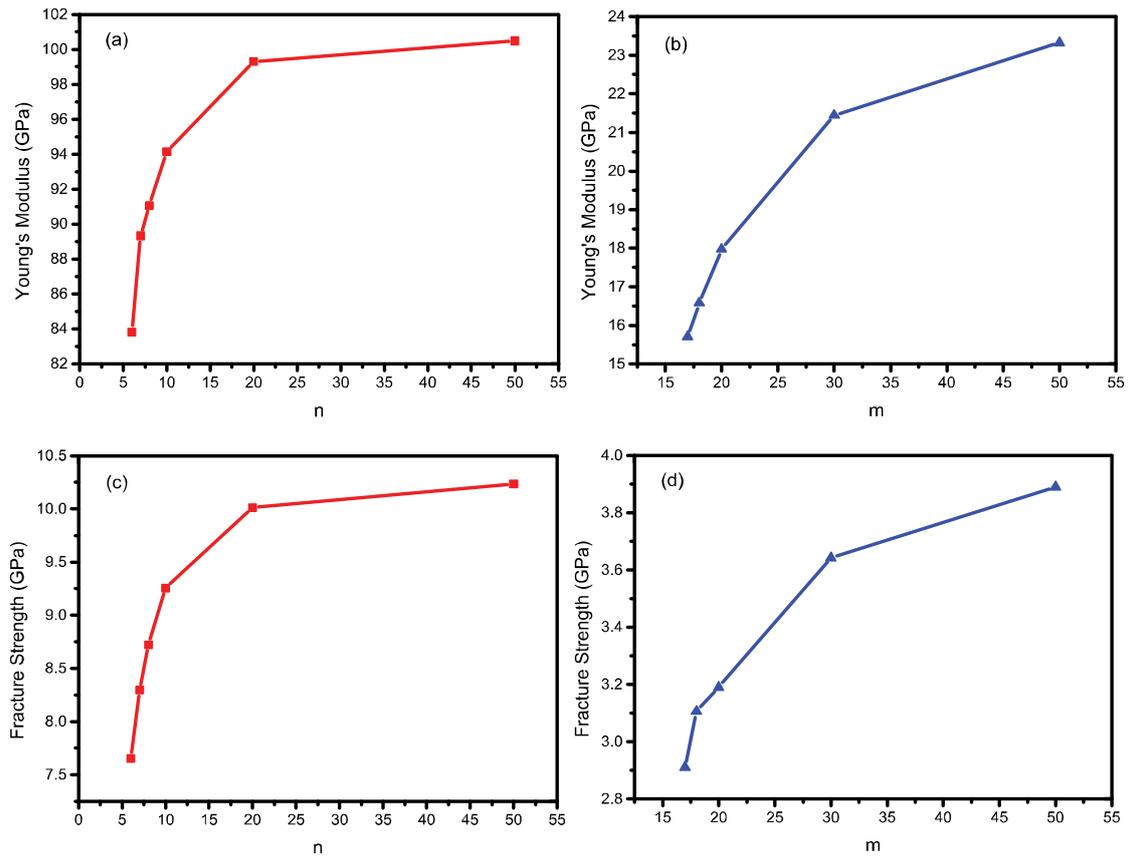

**Figure 4**. Size-dependent mechanical properties at *T* = 0 K. Young's modulus of (a) armPNTs and (b) zigPNTs, fracture strength of (c) armPNTs and (d) zigPNTs.

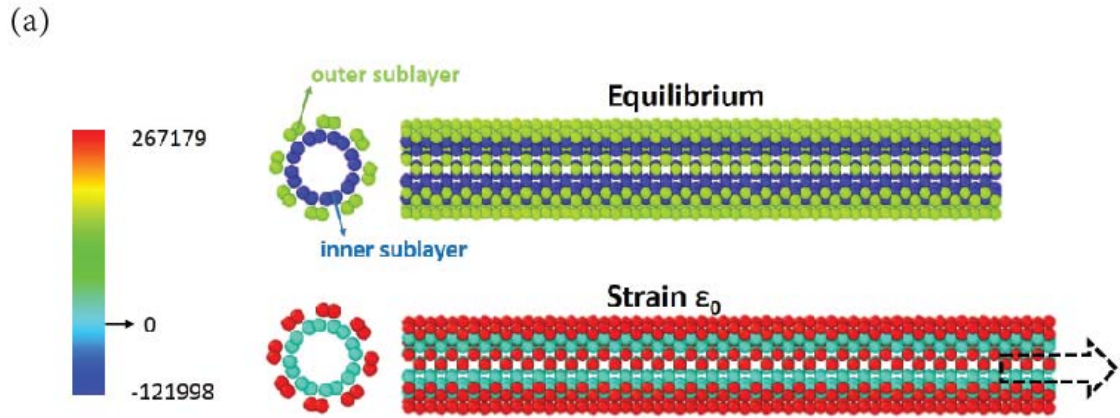

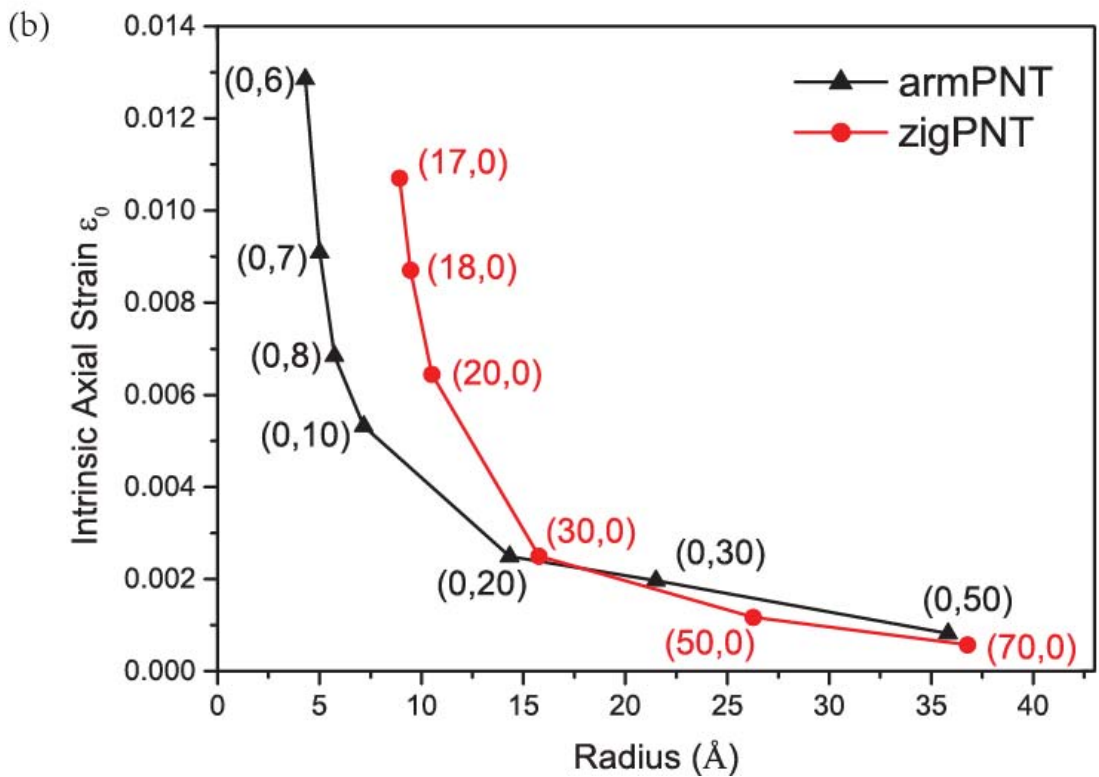

**Figure 5.** (a) The atomic stress (unit: eV/atomic volume) of PNT structures (Top) in equilibrium and (Bottom) under tension $\varepsilon_0$, where the atoms in the inner sublayer are stress-free. (b) The intrinsic axial strain in each sublayer of equilibrium armPNTs (black line) and zigPNTs (red line) at $T = 0$ K.

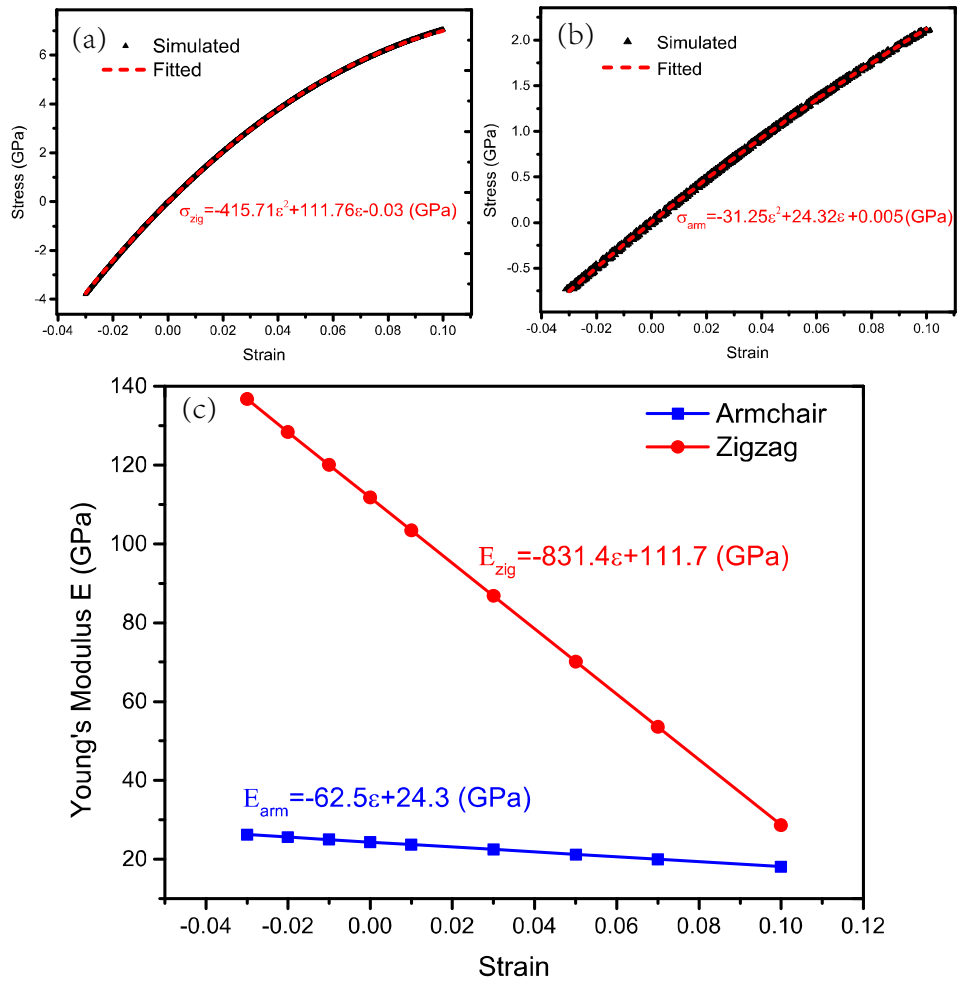

**Figure 6.** The nonlinear stress-strain curves of a 2D phosphorene sheet simulated by MD at $T$ = 0 K along the (a) zigzag direction and (b) armchair direction. Data are fitted by quadratic functions (red dash line) and expressions are shown. Based on the derivatives of the fitting functions, the Young's moduli of the phosphorene sheet along zigzag direction (red line) and armchair direction (blue line) are deduced and plotted as a function of the applied strain, in (c).

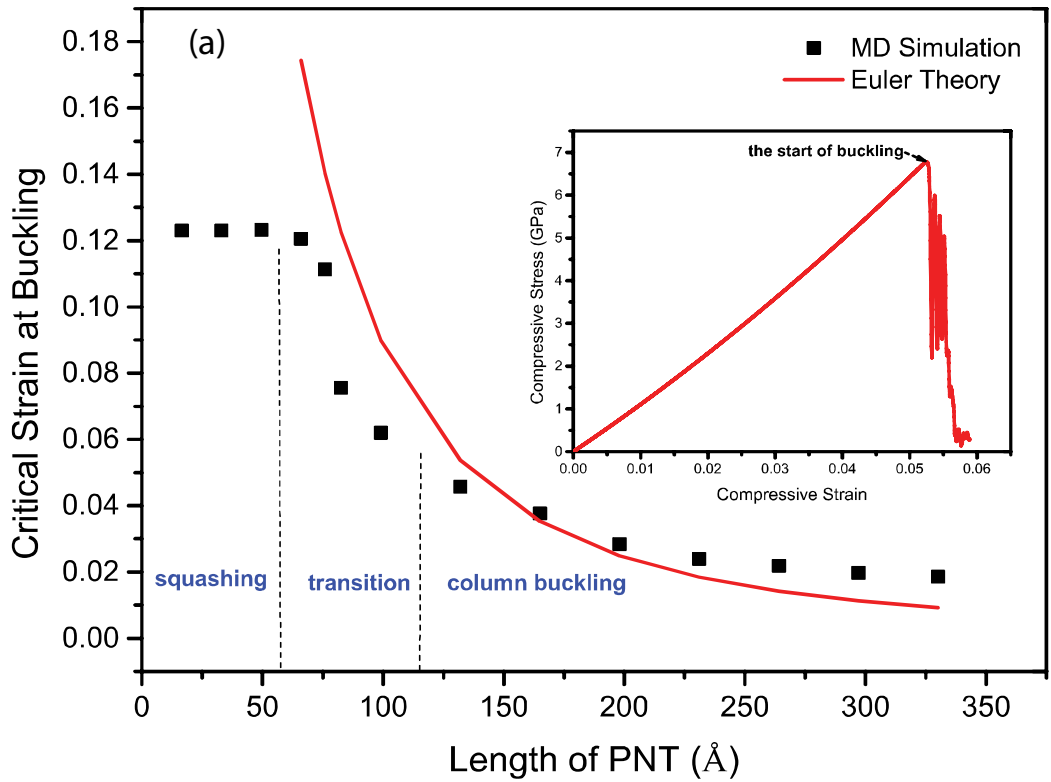

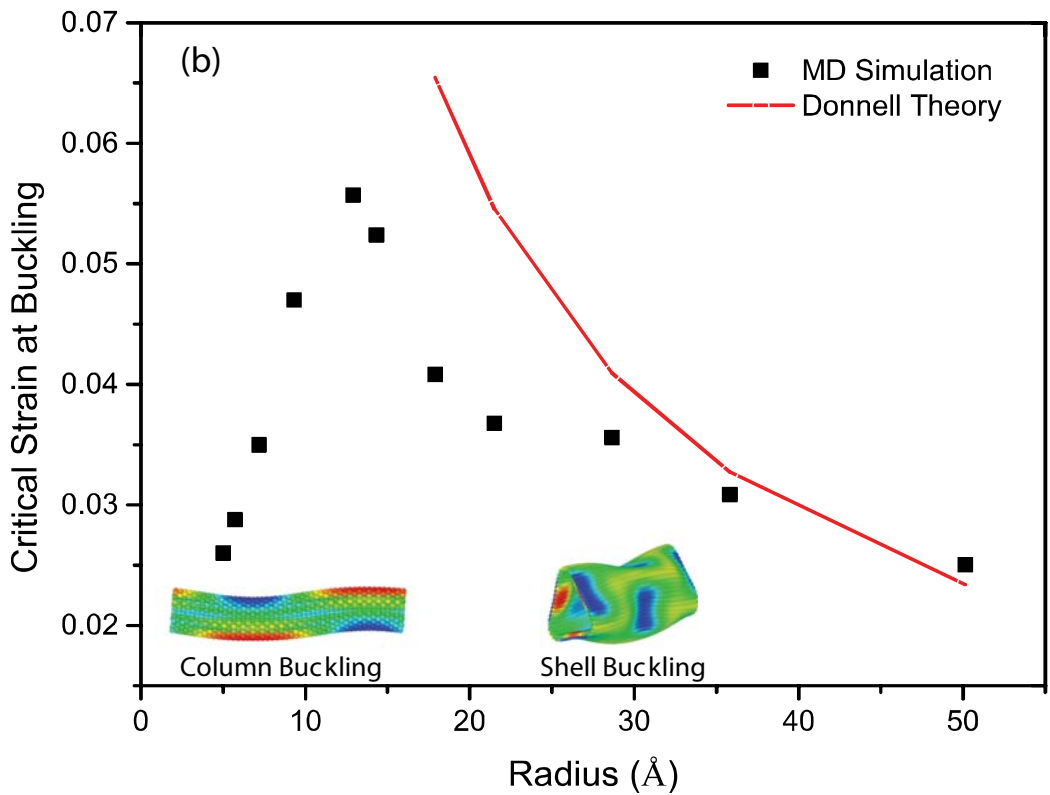

**Figure 7.** The critical strain of buckling versus the geometries of armPNTs. (a) The critical strain vs length for the (0, 10) armPNTs, where the inset shows the definition of buckling strain under axial compressive load. (b) The critical strain vs diameter for the armPNTs with a constant length of 165 Å, showing buckling modes, i.e. the column buckling and shell buckling.